# Impact of substrate on magnetic phase coexistence in bicritical $Sm_{0.53}Sr_{0.47}MnO_3$ thin films

M. K. Srivastava,[a,c] M. P. Singh,[b*] P. K. Siwach,[a] A. Kaur,[c] F. S. Razavi,[b] and H. K. Singh[a#]

[a]National Physical Laboratory (Council of Scientific and Industrial Research), Dr. K. S. Krishnan Marg, New Delhi-110012, India
[b]Department of Physics, Brock University, St. Catharines, Ontario, L2S 3A1, Canada
[c]Department of Physics and Astrophysics, University of Delhi, Delhi-110007, India

**Abstract**

$Sm_{0.53}Sr_{0.47}MnO_3$ thin films were deposited on single crystal $LaAlO_3$ (LAO/(001)) and $SrTiO_3$ (STO/(001)) substrates by DC magnetron sputtering. The θ-2θ and ω-2θ scans show that these films have very good crystallinity and the films on LAO and STO are under compressive and tensile strain, respectively. The films on LAO and STO substrates show ferromagnetic (insulator-metal) transition at $T_C$~126 K (at $T_{IM}$~128 K) and 120 K ($T_{IM}$~117K), respectively. The magnetic state at $T<T_C$ is akin to cluster glass, which is formed by the presence of charge ordered-antiferromagnetic clusters in the ferromagnetic matrix. Huge drop in the resistivity at $T_{IM}$ and the associated hysteresis with respect to cooling and warming cycles reveal the bicritical and the first order nature of phase transition, which is also confirmed by the Banerjee criterion. The differences and similarities in the functional properties of films are explained in terms of substrate modified magnetic phase coexistence.

## 1. Introduction

Among manganites $Sm_{1-x}Sr_xMnO_3$ is unique due to its proximity to the charge order/orbital order (CO/OO) instability and shows the most abrupt insulator metal transition (IMT) [1-3]. The ground states of the low bandwidth (W) $Sm_{1-x}Sr_xMnO_3$ are (a) ferromagnetic metallic (FMM) at $0.3<x\leq0.52$, and (b) antiferromagnetic insulating (AFMI) for $x >0.52$ [1-3].The charge ordering (CO) occurs in the range $0.4\leq x\leq0.6$ and the corresponding ordering temperature ($T_{CO}$) increases from ~140 to 205 K as $x$ increases in the above range. Colossal magnetoresistance (CMR) is observed at all the compositions corresponding to the FMM ground state. Near half doping ($0.45\leq x\leq0.52$), very sharp (first order) transitions from paramagnetic insulating (PMI) to the FMM state are observed. Such abrupt first order PMI-FMM transition is one of the prerequisites of a material having large magnetocaloric effect (MCE) and hence is of technological interest [4,5]. Multiplicity of magnetic phases at $x=0.45-0.52$ provides natural tendencies towards phase separation/coexistence that make the $x$–T phase diagram fragile vis-a-vis external perturbations. Hence, even mild external stimuli such as electromagnetic field, pressure, and substrate induced strain can dramatically modify their physical properties [1-3].

$Sm_{1-x}Sr_xMnO_3$ having a poly- and single- crystalline bulk forms have been studied extensively [2-7]. Very few reports [8-10] exist in literature on single crystalline thin films. One of the possible reasons being the difficulty in growth of thin films with well controlled behaviors, due to the extreme sensitivity of the electronic phases to the substrate induced pseudomorphic strains. It is known that the compressive strain favours FMM and is inimical to the AFM-COI phases while the tensile strain enhances the later at the cost of the former [11]. However, thin films even on (001) $LaAlO_3$ (LAO) substrate which provides compressive strain, have not been reported in literature so far. In this communication we report the magnetic and magnetotransport properties of single crystalline epitaxial thin films of $Sm_{0.53}Sr_{0.47}MnO_3$ (SSMO) grown on (001) LAO and $SrTiO_3$ (STO) substrates. Our studies demonstrate that the functional properties of these films are extremely sensitive to substrate induced strain effect.

## 2. Experimental

Films with thicknesses of about 200 nm were deposited by on-axis dc magnetron sputtering [12] of 2″ $Sm_{0.53}Sr_{0.47}MnO_3$ target on the single-crystal LAO and STO substrates maintained at 800 °C at a dynamic pressure of 200 mtorr of Ar (80%) + $O_2$ (20%). The as grown films were found to be insulating and therefore were annealed in flowing oxygen at 900°C for 12 hrs.

Shorter annealing time resulted in poor characteristics. The structure/microstructure was characterized by powder X-ray diffraction (θ - 2θ and ω-2θ scans) and the cationic composition was studied by energy dispersive spectroscopy (EDS). Temperature and magnetic field dependent magnetization was measured by using a commercial SQUID magnetometer (MPMS-Quantum Design) and the electrical transport was measured by standard four probe method in the temperature range 4.2-300 K.

## 3. Results and Discussion

The cationic stoichiometry of these films was verified by EDS and the same was found to be in very good agreement with that of the target. The lattice constants of the target material are a = 5.448 Å, b = 5.433 Å, and c = 7.672 Å. This yields the average in-plane and out-of-plane cubic lattice constants, $a_c^{in}$ = 3.847 Å and $a_c^{out}$ = 3.836 Å, respectively. STO (3.905 Å) has larger, while LAO (3.798 Å) possesses smaller lattice constant than $a_c^{in}$. Hence, films on STO and LAO are expected to grow under tensile strain (ε = 1.49 %) and compressive strain (ε = -1.29 %), respectively. The XRD (θ - 2θ) data of the SSMO films on LAO (L47) and STO (S47) is plotted (indexed based on the pseudo cubic unit cell) in Fig. 1. The growth in both the films is along the out of plane direction and $a_c^{out}$ of L47 and S47 estimated from the XRD data are 3.851 Å and 3.833 Å, respectively. The slightly larger (smaller) $a_c^{out}$ of L47 (S47) as compared to that of the target, 3.847 Å (3.836 Å) confirms the presence of small compressive (tensile) strain. The full width at half maximum (FWHM) of the ω-2θ scan was found to be 0.09° and 0.11° for L47 and S47, respectively. The AFM pictures of the as deposited and oxygenated films (on LAO) are shown in Fig. 2. The oxygen annealed films show much improved crystallinity. The island like growth is also seen. As suggested by Biswas et al. [13], such growth is expected to make the distribution of strain (small compressive (tensile) strain in films on LAO (STO)) and create compositional inhomogeneity in the films.

The temperature dependent zero field cooled (ZFC) and field cooled (FC) magnetization data (M(T)) taken at H=100 Oe is shown in Fig. 3. The PM-FM transition temperature ($T_C$) is 126 K and 120 K, respectively for L47 and S47. Further, as we lower the temperature $M_{ZFC}(T)$ shows a cusp like feature at $T_P \approx$ 60 K (L47) and 55 K (S47), and then drops sharply in both the films below this point. In the $M_{FC}(T)$ curves, $T_P$ is shifted to lower temperature and the sharpness of the magnetization drop at $T<T_P$ is reduced considerably. The $M_{ZFC}(T)$ and $M_{FC}(T)$ curves of both the films show strong divergence at $T<T_C$. The M-H data (taken at 10 K) of these films is presented in the inset of Fig. 3. L47 (S47) shows saturation magnetization ($M_S$) ≈ 466 emu/cm$^3$ (440 emu/cm$^3$) at a saturation field of $H_S$ ≈ 6 kOe (9 kOe), and remnant magnetization ($M_r$) ≈ 258 emu/cm$^3$ (242 emu/cm$^3$), respectively. Generally, $M_{ZFC}(T)/M_{FC}(T)$ divergence, the cusp-like behavior and the drop in ZFC magnetization at $T<T_P$ are regarded as signatures of metamagnetic state [16]. In the present case, the occurrence of (i) strong ZFC-FC divergence just below $T_C$, (ii) temperature dependence of FC magnetization at $T<T_P$, and (iii) shifting of the $T_P$ towards lower temperature in the FC curve, suggest that the magnetic state in the lower temperature regime is akin to cluster glass (CG) rather than spin glass (SG) [14,15]. The CG appears at the ZFC-FC divergence point, viz., $T_{CG}$ ≈ 116 K and 112 K in L47 and S47, respectively. The sharp drop in $M_{ZFC}$ is a generic signature of cluster freezing at temperature $T_f$, which is ≈36 K and 26 K for L47 and S47, respectively. Interestingly at $T>T_f$, the ZFC-FC divergence is stronger in the compressively strained film (L47), while at $T<T_f$, the same is more pronounced in the film with small tensile strain (S47). The occurrence and coexistence of the short range CO-AFMI and A-AFM correlations at $T>T_C$ in bulk $Sm_{1-x}Sr_xMnO_3$ ($x$~0.5) is well established [1,2]. However, in thin films the structural-microstructural modifications induced by strain could lead to appreciable change in the magnetic landscape even at $T<T_C$. Hence, in the present case (i) long range FMM, (ii) short range CO/OO-AFMI phases and (iii) short range A-AFM phases are expected to coexist also at $T<T_C$. Such phase-coexistence results in for formation of the CG at $T<T_C$.

The lower $T_C/T_{IM}$ in S47 is due to the presence of tensile strain that favours CO-AFM rather than FM ones [11]. However, it is interesting to note that at smaller magnetic fields (<500 Oe) the magnetic moment of L47 are smaller than that of S47 and the reverse is true at higher fields. The observed difference in magnetic moments of the two set of films could possibly be due to the different nature of CG in the two set of films. This could arise from disorder caused by the difference in the oxygen stoichiometry and different nature of substrate induced strain in these films. The CG is known to be sensitive to disorder [15]. In the present case disorder could arise due to the different impacts of the oxygen deficient and magnetically disordered dead layer in compressively strained (L47) and tensile strained (S47) films [16]. The higher values $T_f$ in films on LAO than that on the STO substrates could also be attributed to the slightly different nature of the CG state in the two set of films.

The strong sensitivity of magnetization to the small magnetic fields coupled with the sharp PM-FM transition suggests first order phase transition (FOPT). To verify the nature of the magnetic transition we measured the isothermal magnetization in the vicinity of the PM-FM transition. The nature of the magnetic phase transition was evaluated from Banerjee criterion by plotting $M^2$-H/M data [17]. The representative $M^2$-H/M data is plotted in Fig. 4. The change in the slope of the $M^2$-H/M curve of both (L47 and S47) films from negative around $T_C$ to positive at $T<T_C$ shows that the PM-FM transition is of the first order.

To further evaluate the nature of magnetic phases in these films, we analyzed the difference between the $M_{ZFC}(T)$ and $M_{FC}(T)$ magnetization ($\Delta M = M_{ZFC} - M_{FC}$) of the two films. As shown in the inset of Fig. 4, $d(\Delta M)/dT$ starts increasing at $T_{CG}$ in both L47 and S47 and the rise is appreciably sharper in the former. As the temperature is reduced further, a plateau like region appears, the minimum of which corresponds $T_P$. In the lower temperature region, $d(\Delta M)/dT$ shows a peak and the temperature corresponding to this peak is the cluster freezing temperature $T_f$. The considerably sharper peak in the films grown on STO substrate suggests stronger cluster freezing. The noted difference in the temperature dependence of $d(\Delta M)/dT$ of these films is caused by the difference in the dimensionality, density and fraction of the competing magnetic phases. As mentioned above, STO provides tensile strain that is friendly (inimical) to the CO-AFMI (FMM). Hence in the films on STO, concentration and size of the CO-AFMI phase is expected to get enhanced at the cost of FMM even at $T<T_C$.

Temperature dependence of resistivity ($\rho$-T) measured in the range 4.2-300 K using the protocol of slow cooling and subsequent warming in zero magnetic field is shown in Fig. 5. At the IMT $\rho(T)$ drops by nearly four orders of magnitude within a very small temperature range. During the warming cycle $\rho(T)$ grows irreversibly shifting the IMT to higher temperature. This hysteretic behaviour of $\rho(T)$ coupled with the huge drop in resistivity is a generic feature of the first order transition observed in low W manganites and has been reported for single crystalline $Sm_{1-x}Sr_xMnO_3$ ($0.45 \leq x \leq 0.52$) [1,2]. The IMT occurs at 117 K (112 K) and 128 K (118 K) during cooling and warming cycles in L47 (S47), respectively. Interestingly, in the films on LAO, $T_C < T_{IM}$, while the reverse is true for those on STO. At $T > T_{IM}$, $\rho(T)$ of both L47 and S47 is nearly similar. However, $\rho(T)$ of L47 and S47 at $T < T_{IM}$ show contrasting characteristics e.g., the latter has a hump like feature in the range T~80 - 100 K that shifts to slightly higher temperature in the warming cycle. As discussed above, this could be due to the enhanced CO-AFMI fraction in the films on STO. The hump like feature in the films on STO is very sensitive to oxygen annealing and disappears when these films are subjected to prolonged oxygen annealing. Application of even a small magnetic field suppresses the hump and this causes the second MR peak in the films on STO. The sharpness of the IMT is evaluated by the temperature coefficient of resistivity (TCR) defined by $d(\ln\rho)/dT$. In L47, the peak value of TCR~150 % (115 %) occurs at T~107 K (115 K) in the cooling (warming) cycle. S47 shows two peaks in TCR; one appears just below the $T_{IM}$ at 98 K (110 K) in the cooling (warming) cycle. The second TCR peak occurs in the region of the resistivity hump at 81 K (85 K) in the cooling (warming) cycle. The TCR measured in the warming cycles is plotted in the upper inset of Fig. 5.

The relative variation in the $T_C$ with respect to $T_{IM}$ in these films could be cause by the presence of spatially non-uniform strains and compositional inhomogeneity as a consequence of the island type growth [13]. The $T_C>T_{IM}$ in films on LAO suggests that metallic fraction reaches the percolative threshold earlier than the FM fraction. This may be suggestive of the presence of small fraction of the metallic A-AFM phase in films on LAO. Further, $T_C<T_{IM}$ in films on STO could be due to the presence of some insulating FM phase, so that FM fraction approaches the percolative threshold earlier than the metallic fraction. Since the relative magnitudes of $T_C/T_{IM}$ are sensitive to type of strain, we suggests that strain induced effects are responsible for the observed effect.

Both the films show huge low field magnetoresistance (LFMR, measured at H = 3 kOe). As shown in the lower inset of Fig. 4, L47 shows a peak LFMR $\approx$55 % at 120 K. In S47, like its TCR, two LFMR peaks are observed, the first at 112 K (MR$\approx$40%) and the second at 98 K (MR$\approx$64 %). The occurrence of the lower MR peak in S47 is due to the suppression of the hump resistivity by the magnetic field and the larger value of MR is indicative of enhanced phase fluctuation.

It has been reported earlier that in $Sm_{1-x}Sr_xMnO_3$ (0.45$\leq x \leq$0.52) coexisting FMM and AFMI phases and their dimensionality that determine the magneto-electric landscape are critically sensitive to external perturbations and quenched disorder [1,2,6,7,15]. Around the room temperature, SSMO is PMI; however, nanoclusters of CO-AFMI and FMM phases may be embedded therein [1,2]. On lowering the temperature the fraction of the CO-AFMI is enhanced. Just above $T_C/T_{IM}$ the fraction of the CO-AFMI insulating phase rises rapidly resulting in sharp enhancement in the resistivity. At $T_C/T_{IM}$, CO-AFMI and FMM coexistence is delicately balanced resulting bicriticality and any external perturbation/stimuli (e.g., temperature and magnetic field) can easily destroy the unfavorable phase. Consequently, the CO-AFMI phase is suddenly removed (the FMM phase appears) when the temperature is even slightly lower than the $T_{IM}/T_C$ resulting in the sharp drop in resistivity. However, the short range CO-AFMI clusters could still be present at $T < T_C$ and may cause the occurrence of CG, as described above. The irreversible M(T) and the cooling – warming hysteresis in the $\rho(T)$ data are attributed to the slow evolution of the phase conversions among coexisting phases in the material.

The hysteresis in the resistivity and concomitant difference in the IMTs could be understood in terms of the spin CG concept. During the cooling cycle the spin CG mimics a liquid like behavior and the carrier scattering by disordered spins is strong. This results in large resistivity and lower $T_{IM}$. When cooled below $T<T_f$, clusters are frozen and in this state the carrier scattering is considerably suppressed leading to the lower resistivity and higher $T_{IM}$ during the warming cycle. The occurrence of the hump and the corresponding double peaks in TCR and LFMR could possibly be due to the re-enforcement of the CO-AFMI phase due to oxygen deficiency and inhomogeneous strain. The films on STO are, as discussed earlier, under tensile strain, which favors the CO-AFMI phase and is inimical to the FMM [11]. This enhances magnetic phase fluctuations that increases the carrier scattering resulting in (i) $T_C>T_{IM}$, (ii) relatively broad IMT, (iii) the observed hump like feature in the $\rho(T)$ curve at $T<T_{IM}$ and (iv) the lower peak in the TCR and MR. When a magnetic field is applied the magnetic-spin fluctuation is suppressed strongly resulting in larger LFMR in the vicinity of the resistivity hump.

## 4. Conclusion

We have prepared single crystalline $Sm_{0.53}Sr_{0.47}MnO_3$ thin films on LAO and STO substrates and studied the modification in the structure, magnetic and transport properties. Our study shows that the compressive strain favours the FMM, while even a very subtle tensile strain favours AFM-COI phases. Both the films show first order phase transitions, which is confirmed by the negative slope of the $M^2 - H/M$ curves in the vicinity of $T_C$ and a very sharp and hysteretic IMT. The presence of AFM-COI phases in the FMM matrix results formation of a cluster glass like state. Both set of films exhibit huge magnetoresistance (MR~60 %) even at small magnetic fields of ~3 kOe. In the films on STO, which are under small tensile strain a strong phase fluctuation at $T<T_C/T_{IM}$ results in a hump in the resistivity below $T_{IM}$ that in turn causes second peak in the MR and TCR.

**Acknowledgements**

MKS is thankful to CSIR, New Delhi for junior research fellowship. Authors at NPLI acknowledge the continued support from Prof. R. C. Budhani. At St Catharines, this work was supported by NSERC (Canada), CFI, Ontario Ministry of Research and Innovation (MRI), and the Brock University.

**References**

1. Y. Tokura, Rep. Prog. Phys. 69 (2006) 797.
2. Y. Tomioka, H. Hiraka, Y. Endoh, and Y. Tokura, Phys. Rev. B 74 (2006) 104420
3. C. Martin, A. Maignan, M. Hervieu, and B. Raveau, Phys. Rev. B 60 (1999)12191.
4. A. Rebello and R. Mahendiran, Appl. Phys. Lett. 93 (2008) 232501.
5. P. Sarkar, P. Mandal, and P. Choudhury, Appl. Phys. Lett. 92 (2008)182506.
6. M. Egilmez, K. H. Chow, J. Jung, and Z. Salman, Appl. Phys. Lett. 90 (2007) 162508.
7. M. Egilmez, K. H. Chow, J. Jung, I. Fan, A. I. Mansour, and Z. Salman, Appl. Phys. Lett. 92 (2008) 132505.
8. H. Oshima, K. Miyano, Y. Konishi, M. Kawasaki, and Y. Tokura, Appl. Phys. Lett. 75 (1999) 1473.
9. M. Kasai, H. Kuwahara, Y. Tomioka, and Y. Tokura, J. Appl. Phys. 80 (1996) 6894.
10. M. Egilmez, M. Abdelhadi, Z. Salman, K. H. Chow, and J. Jung, Appl. Phys. Lett. 95 (2009) 112505.
11. W. Prellier, Ph. Lecoeur and B. Mercey, J. Phys. Cond. Matter 13 (2001) R915.
12. R. Prasad, M. P. Singh, P. K. Siwach, W. Prellier, and H. K. Singh, Solid State Commun. 142 (2007) 445.
13. Amlan Biswas, M. Rajeswari, R. C. Srivastava, Y. H. Li, T. Venkatesan, R. L. Greene, and A. J. Millis, Phys. Rev B 61 (2000) 9665.
14. J. A. Mydosh, Spin Glasses: An Experimental Introduction, 2nd ed. (Taylor & Francis, London, 1993).
15. M. K. Srivastava, P. K. Siwach, A. Kaur, and H. K. Singh, Appl. Phys. Lett. 97 (2010) 182503.
16. J. Z. Sun, D. W. Abraham, R. A. Rao, and C. B. Eom, Appl. Phys. Lett. 74 (1999) 3017.
17. S. K. Banerjee, Phys. Lett. A12 (1964) 16.

**Figure Captions**

Fig. 1. X-ray diffraction patterns (θ-2θ scan) of L47 (LAO) and S47 (STO) thin films. Substrate peaks are marked S and SSMO reflections are indexed.

Fig. 2. Representative AFM images of (a) the as grown film, and (b) oxygen annealed film on LAO.

Fig. 3. Temperature dependent ZFC & FC magnetization (@H=100 Oe). Various transitions are marked by arrows. Inset shows the M-H loops measured at 10 K.

Fig. 4. Isothermal $M^2$ - H/M plots at different temperatures for L47 and S47. Inset shows temperature dependence of d(ΔM)/dT.

Fig. 5. Temperature dependence of resistivity measured in the range 4.2-300 K during cooling (marked by C) and warming (W) cycles. The arrow indicates the hump in the resistivity. Lower and upper insets show the LFMR and TCR, respectively measured during the warming cycle.

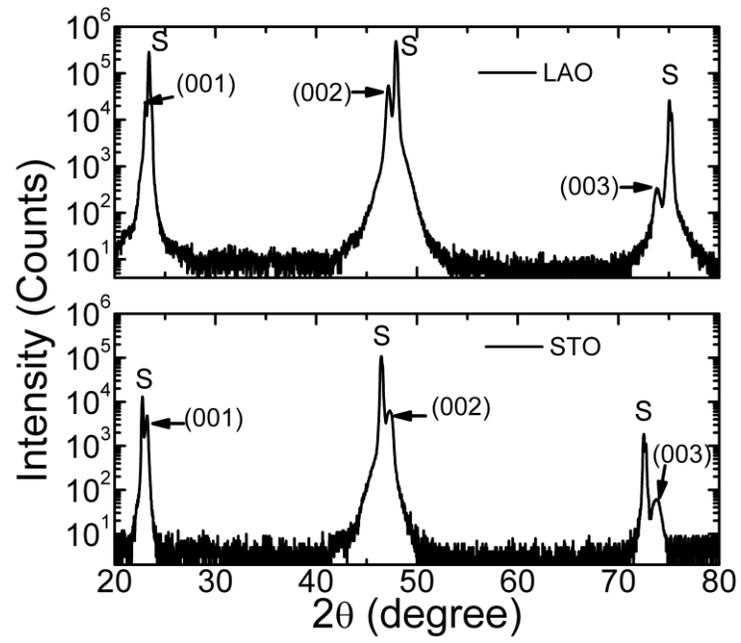

Fig. 1

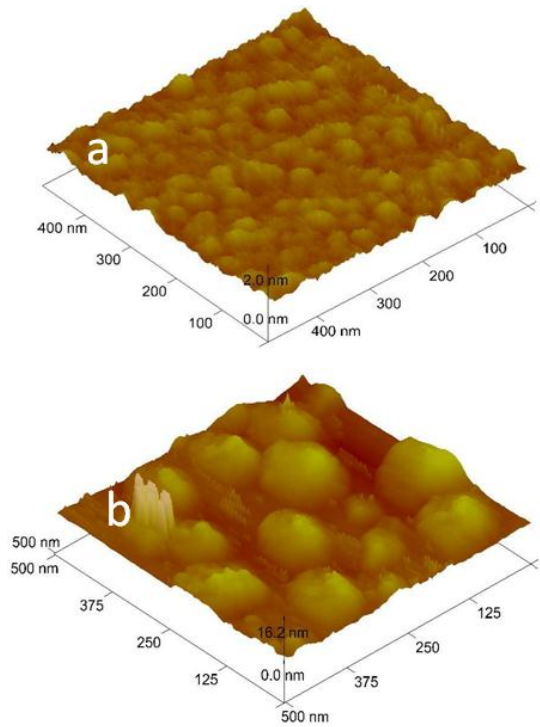

Fig. 2

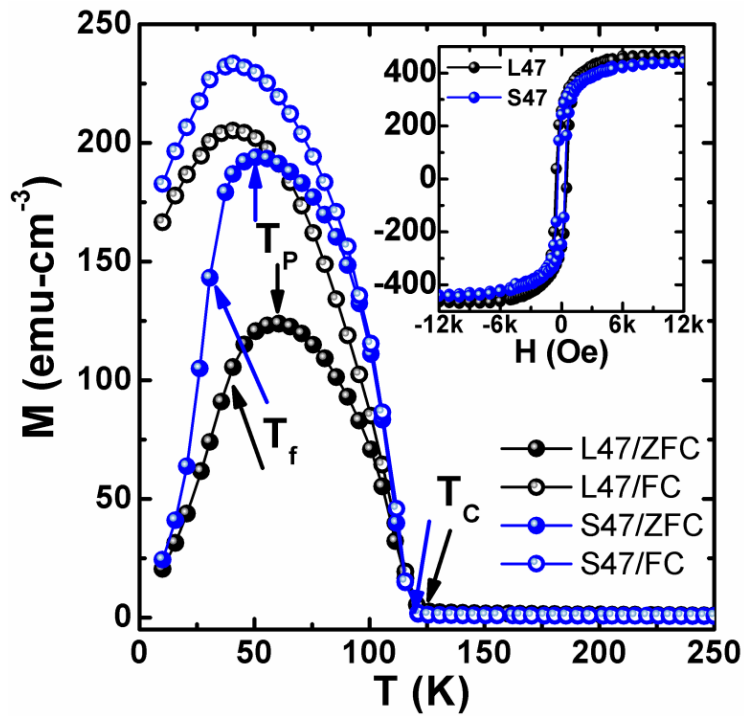

Fig. 3

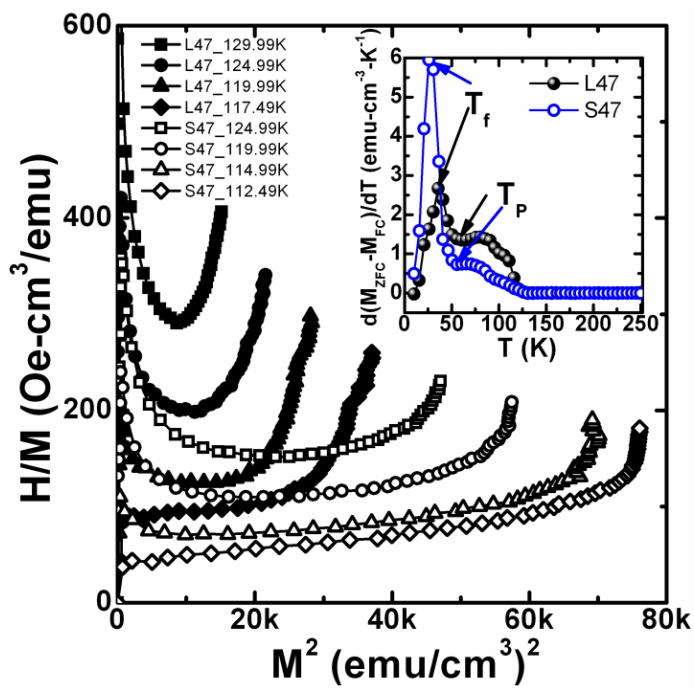

Fig. 4

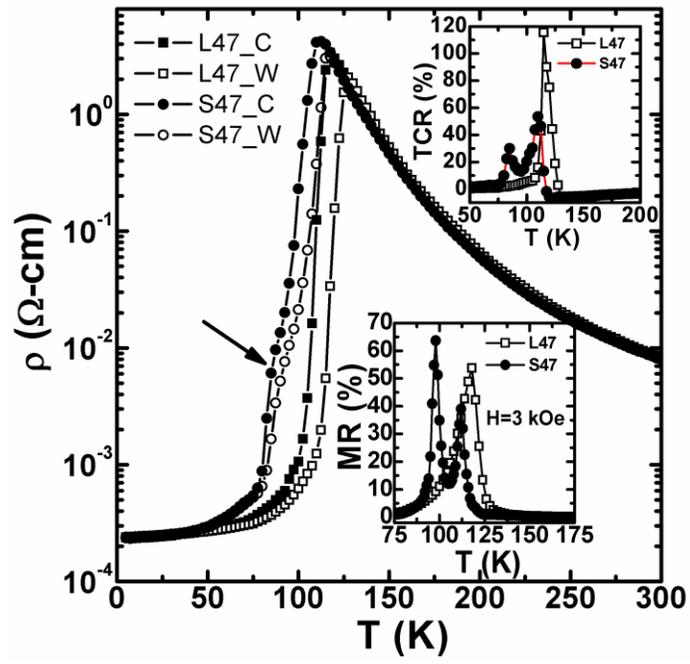

Fig. 5